# Recent Advances in Computational Modeling of Thrombosis

Sumith Yesudasan and Rodney D. Averett*

School of Chemical, Materials and Biomedical Engineering, University of Georgia
597 D.W. Brooks Drive, Athens, GA 30602

*Corresponding author email: raverett@uga.edu
Telephone: 706-542-0863

## Abstract

The study of thrombosis is crucial to understand and develop new therapies for diseases like deep vein thrombosis, diabetes related strokes, pulmonary embolism etc. The last two decades have seen an exponential growth in studies related to the blood clot formation using computational tools and through experiments. Despite of this growth, the complete mechanism behind thrombus formation and hemostasis is not known yet. The computational models and methods used in this context are diversified into different spatiotemporal scales, yet there is no single model which can predict both physiological and mechanical properties of the blood clots. In this review, we will attempt to list out all major strategies attempted by researchers so far to model the blood clot formation using existing computational techniques. This review classifies them into continuum level, system level, discrete particles and multi-scale methods. We will also discuss the strength and weakness of various methods and possible future directions in which the computational blood clot research can thrive.

## Introduction

Hemostasis is a natural process of stopping the excessive bleeding in an event of injury[1]. On the other hand, Thrombosis is often related with pathological states and is the formation of blood clots (thrombus) inside the vascular chamber due to endothelial damage or other causative events, obstructing the flow of blood. Thrombosis can often lead to the rupturing of the thrombus, forming embolisms which is a leading cause for the arterial embolism, stroke, myocardial infarction, venous thrombo embolism[2] (VTE) and pulmonary embolism.

Despite the advancements in the blood clot studies by researchers, our understanding about the blood clotting process is incomplete and still evolving[3-4]. The knowledge about the mechanics behind these events is of significant importance to develop more effective thrombolytic therapies or to prevent it. In comparison with physical experiments, computational modeling can give insights at nanoscale properties of biochemical processes which may affect the mechanical properties at macroscale. For example, studies like estimation of fibrinogen elasticity from molecular simulations[5] can be scaled up to macroscale to understand the bulk behavior. The overwhelming amount of literature in thrombosis experiments point outs the risk factors associated with the VTE formation but the root cause under various pathological conditions are not yet identified. To unveil the reason behind such diseases, we need to have insights about the relationship of thrombus formation and rupture with its dynamic mechanical properties. This calls for the understanding of molecular basis and mechanisms by which the clot mechanical properties are governed. The developments in computational methods like molecular simulations, mesoscopic methods and continuum theories promise a synergetic solution to these kind of problems. Towards achieving this, we will summarize the advancements in the area of blood clot computational modeling and simulation of thrombus formation.

Past two decades show an exponential increase in number of publications related to understanding the formation of blood clots under various circumstances and physical locations. Figure 1 shows the results of *web of knowledge* search results of number of publications related to computational modeling of thrombus formation and experimental investigation of blood clot formation mechanisms. This shows the growing field towards the investigation of blood clot formation and its computational modeling. The studies range from molecular level to large scale continuum level and from blood constituents like fibrinogen[6], erythrocytes, platelets, hemoglobin[7], fibrin mechanics[8] etc. A short review on this field is given by [9]. Also a review on discrete particle methods used for blood simulation is given by [10].

Despite of the advancements in experimental investigation of blood clotting process, computational modeling and theoretical understanding of the same is not well developed. In this review, we will consolidate the computational modeling and strategies used by researchers to study the blood clot formation. The emphasis of the review will be given to direct modeling techniques to simulate the clot formation and its dynamic behavior, rather than focusing on the computational methods of experimental imaging [11-12].

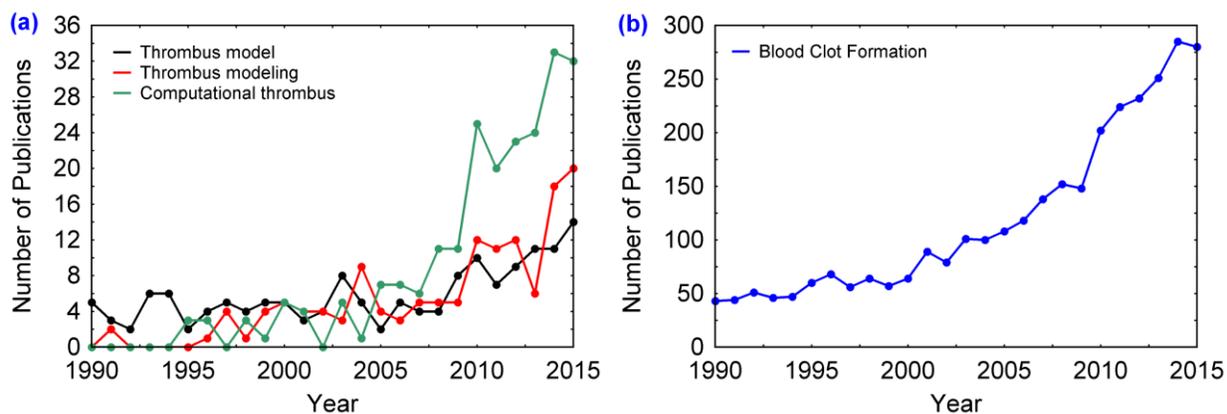

Figure 1. Total number of publications over the past 15 years with different keywords related to the thrombus formation modeling. (a) Computational studies related to the thrombus formation (b) Publications related to the study of blood clot formation.

## Biological Background

Hemostasis[13] is a natural process to keep blood within a damaged blood vessel to prevent hemorrhage[1]. When endothelial injury occurs, the endothelial cells stop secretion of coagulation and aggregation inhibitors and instead secrete von Willebrand factor[14] which initiate the maintenance of hemostasis after injury. Hemostasis has three major steps: 1) vasoconstriction, 2) temporary blockage of an injury by a platelet plug, and 3) blood coagulation, or formation of a fibrin clot[15]. These series of processes seal the hole until tissues are repaired. Blood coagulation step consists of a cascade of events in which a number of blood factors participate in biochemical reactions leading to the formation of a stable clot. This is explained briefly in the next section. In the normal condition, when there is no injury, the endothelial cells of intact (healthy) vessels prevent blood clotting with production of prostacyclin, secretion of plasminogen activator, cell surface heparin containing proteoglycans that stimulate certain coagulation factor inhibitor, and thrombomodulin, an endothelial cell surface glycoprotein that promotes protein C activation[16].

**Coagulation Cascade**

The classical coagulation cascade[17] consists of intrinsic path, extrinsic path and a common pathway. Over the past few decades, our understanding about the coagulation cascade is changed and is continually improving[17-19]. The modern coagulation cascade consists of an i) Initiation phase, ii) Amplification phase, and iii) Propagation phase.

The initiation phase starts with the exposure of the tissue factor (TF), present in the sub-endothelial cells with blood stream during an event of vascular injury. The TF binds with factor VIIa (FVIIa) and cleaves FIX and FX to form FIXa and FXa respectively. The suffix *a* after the blood factors indicate that they are in the activated state and the prefix *F* stands for factor. FXa will then help conversion of FII to FIIa (thrombin). This is followed by the amplification phase, in which the FIIa plays a central role as shown in the Fig. 2. FIIa will convert FV to FVa, FXI to FXIa and FVIII to FVIIIa. These parallel processes will speed up the production of FIIa and is hence called amplification phase. In the propagation phase, the FXIa also converts FIX into FIXa, which along with FVIIIa catalyzes the formation of FXa. Finally, FXIIIa catalyzes the formation of cross linked fibrin network. A detailed explanation of these steps and other sub processes can be found in the literature[18] and a brief review about the fibrinolysis is provided by Longstaff[20].

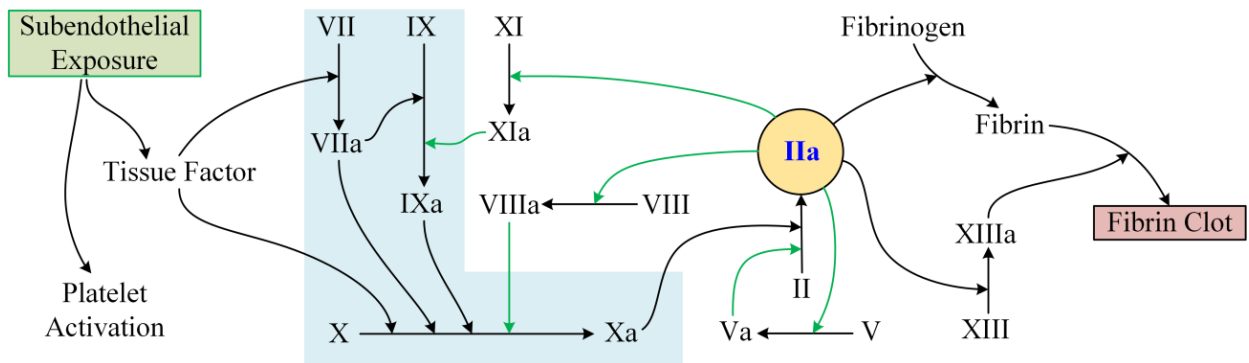

Figure 2. Modern coagulation cascade. The light blue shade indicates the initiation phase and green arrows show amplification phase. The blood factors and their names are listed in the nomenclature section. Note that this cascade doesn't show clot inhibiting factors like protein C etc.

**Thrombosis and Thrombus Formation**

Thrombosis is the process of formation of a blood clot inside a blood vessel, obstructing the flow of blood through the circulatory system, which is different from the natural process hemostasis. Even when a blood vessel is not injured, blood clots may form in the body under certain conditions and is mostly related to pathological states. A clot, or a piece of the clot, that breaks free and begins to travel around the body is known as an embolus. In platelet poor conditions as in veins, emboli can travel to lungs and form pulmonary embolism, or can cause a number of other conditions like deep vein thrombosis, renal vein thrombosis, etc. In the case of arterial thrombosis, the emboli can lead to stroke, or myocardial infarction. Insights about the mechanical properties of the blood clot can be helpful to develop advanced thrombolytic therapies[21]. In their study, Weisel have observed that fibrin clots are made with interconnecting fibrin with numerous branches and crosslinking. However, these long fiber strands do not exhibit any loose ends and are viscoelastic. They have also examined the effects of XIIIa on viscoelastic properties, strain hardening, sensitivity of viscoelastic properties to small changes in polymerization and network strands dissociation[21]. In a different study, Ryan[22] suggested that fibrin diametrical growth is directly related to the branching of the fibrin, and its chances reduces with increased number of branching. The fibrin network stiffness is related to the increased fibrinogen concentration, thickness of the fiber and branch point frequency. Their study also suggest that FXIIIa induced ligation increases the rigidity and stability of the clot.

Due to the large number of biochemical processes, influence of flow conditions on fibrin aggregation and environmental factors, study of thrombus formation, its growth and rupture is challenging. Experimental investigations are often limited or focused to a single or a handful of propertied at a time. For example many of them are focused on rheological properties under various compositions and pathological states, rate of coagulation, effect of specific enzymes or effect of mechanical factors like implants etc. In these situations, computational tools can be very insightful in studying multiple processes and components simultaneously. However, we don't have many computational studies revealing the molecular basis for the mechanical properties as well as the dynamics of clot formation using a single approach.

The processes involved in the clotting process can be broadly classified based on the length scales. The associated enzymatic proteolysis reactions and their associations and dissociations due to the hydrophobic and electrostatic interactions can be treated as molecular level processes, dynamic clot formation and lysis of the larger particles like platelets, vWF, fibrin, RBC etc. can be treated as microscopic processes. The top level polymerization of fibrin, its attachment and detachment with vascular chamber walls and RBCs can be treated as macroscopic continuum level processes. In order to simulate these processes at different spatiotemporal scales, we have to use appropriate computational tools. Luckily, the mathematical and computational methods have advanced very much in the last century and the computational resources like computer and servers have become cheaper and faster. One can make use of the free open source software and powerful languages like LAMMPS[23], NAMD[24], GROMACS[25], OpenFoam[26], Octave[27], Python[28] etc. to simulate these processes at different scales.

## Mathematical Background

To make a choice among the number of tools and methods available for any specific problem, one has to be cognizant both in the problem definition as well as the tool or method. The first choice will be the most familiar tool, which may or may not the best tool for that problem. The large number of computational studies conducted in the area of blood clot follows the same by using a set of standard methods which we will list out in this section. Our idea is to give a brief introductory concept towards various techniques and the length and time scales in which they are applicable. A detailed demonstration or explanation of such methods are beyond the scope of this review and readers are advised to refer the appropriate cited references.

**Navier-Stokes Equation**

The Navier-Stokes (NS) equation[29] might be one of the highly used mathematical equation today and is used to describe the motion of viscous fluid substances. They are widely used to model the weather, ocean currents, water flow in pipes, combustion, air flow around an aircraft wing etc. Recently NS equations are also used to simulate the blood flow in human body under various physiological conditions. The differential form of the NS equations is shown in Eq. 1, with bold face letters representing vectors.

$$\partial \mathbf{u}/\partial t + (\mathbf{u}.\nabla)\mathbf{u} = \nu\nabla^2\mathbf{u} - \nabla w + \mathbf{g} \qquad (1)$$

Here, $\mathbf{u}$ is the velocity of the fluid, $t$ is time, $\nu$ is the kinematic viscosity, $\mathbf{g}$ is gravity or external body force and $w = P/\rho$, where $P$ is pressure and $\rho$ is the mass density. The left hand side terms of the equation represent local variation of velocity and its convection. The terms on the right hand side represents the diffusion term, pressure or internal source term and gravity along with external sources respectively. NS equations are often solved with the assumption of incompressibility, given mathematically as $\nabla.\mathbf{u} = 0$.

As of today, there exists no exact general solution for NS equations, but we can get analytical solutions for simple geometries with assumptions like incompressibility, laminar flow etc. Also for complex geometries we can numerically differentiate NS equations to arrive at approximate solutions. The solution of the NS equation is velocity field and other interesting properties like stream functions, vorticity, pressure variation etc. When the geometry is simple like two flat plates separated at a distance or a pipe, then the flow between them is called Poiseuille flow[29]. Often, blood flow in narrow vascular chambers are simulated using Poiseuille flow.

**Convection Diffusion Reaction Equations**

Convection diffusion reaction (CDR) equations are analogous to NS equations in appearance, but are used to define the transport of a chemical species, electric current or temperature. These equations are called as advection-diffusion equation, drift diffusion equation[30] or scalar transport equations[31], depending on the context of application. A general form of CDR equation is shown below:

$$\partial c/\partial t = D\nabla^2 c - \mathbf{u}.\nabla c + R \qquad (2)$$

Here, $c$ is the variable of interest, concentration for chemical species transport, temperature for heat transfer etc. $\mathbf{u}$ is the velocity of the medium, $D$ is the mass or thermal diffusivity and $R$ is the source or sink. For a chemical species, $R$ is the chemical reaction creating more or less species of $c$. The first term of the Eq. 2 represents the local variation of the species with time. The second term describes the diffusion of the species based on the concentration difference. The third term describes the convection or advection, which describes the change in species concentration due to the flow of the medium around it. The solution of CDR equation is the variation of concentration $c$ with time and space. This is very useful in the context of simulating chemical reactions coupled with flow.

**Dissipative Particle Dynamics**

Dissipative Particle Dynamics (DPD) is a relatively new stochastic simulation technique for simulating the dynamic and rheological properties of simple and complex fluids[32]. It is often considered as a mesoscopic version of MD simulations capable of simulating larger time and length scales[33]. Due to its simplicity, yet powerful capabilities, DPD technique is being used in simulation many complex fluid systems[34], like fibers in viscous medium, dispersion of nano fluids, nano composites, surfactants etc. to name a few of them.

The DPD method treats the fluid system as a collection of particles called as beads, which interact with each other using soft repulsive potentials. The DPD system is governed by Newton's second law and force acting on an $i^{th}$ bead is the sum of internal and external forces, given by $f_i = f_i^{int} + f_i^{ext}$. Here, $f_i^{ext}$ is the external forces on the system by gravity and other body forces. The internal force is given by the mutual interactions:

$$f_i^{int} = \sum_{j \neq i}(F_{ij}^C + F_{ij}^D + F_{ij}^R) \tag{3}$$

The Eq. 3 consists of a soft repulsive conservative force $F_{ij}^C$, a dissipative force $F_{ij}^D$, and a random force $F_{ij}^R$. This will simulate many properties of the fluid, including its density, diffusivity, surface tension etc. The details of implementation, estimation of variables for a specific system, validation etc. are discussed in detail in[33]. One advantage of DPD simulations is that it can handle non-Newtonian flow properties of a fluid[35] which makes it attractive in the case of blood flow modeling[10].

**Cellular Potts Model**

Cellular Potts model (CPM) is a computational model which can simulate collective behavior of cellular structures. CPM is also known as Extended Potts model and is modeled as an extension to the large-q Potts model[36]. It allows modeling of many biological phenomena, such as cell migration, clustering, and growth by considering environment sensing as well as volume and surface-area constraints [37-38].

**Other Computational Methods**

Apart from the above mentioned methods, there exists many newly developed methods used to study the blood flow recently. Lattice Boltzmann methods (LBM) and Smooth Particle Hydrodynamics (SPH) are among them. LBM is a class of computational fluid dynamics (CFD) in which the flow of the fluid is simulated by solving discrete Boltzmann equation with collision models such as Bhatnagar-Gross-Krook[39] (BGK). LBM has advantages of simulating complex, coupled flow with heat transfer and chemical reactions. SPH is a computational method used for simulating the dynamics of continuum media, such as solid mechanics and fluid flows [40]. SPH has several benefits over traditional grid-based techniques in simulating fluid flow. First, SPH guarantees conservation of mass without extra computation since the particles themselves represent mass. Second, SPH computes pressure from weighted contributions of neighboring particles rather than by solving linear systems of equations. Finally, unlike grid based techniques which must track fluid boundaries, SPH creates a free surface for two-phase interacting fluids directly since the particles represent the denser fluid (usually water) and empty space represents the lighter fluid (usually air). The simplicity in implementation makes it as an attractive choice for fluid simulation.

# Computational Modeling of Blood Clot

Modeling of blood and its clot formation is a complex process[41]. Due to the involvement of a large number of proteolysis reactions and polymerizations and its effect on different length scale and time scale makes it difficult to simulate with a single method. In this section we will classify various methods used for hemostasis and thrombosis based on length, time scale and its class. For brevity, we will limit our discussions related to modeling and simulation of blood or fibrin clot formation and also their mechanics. First, we will list out the simulation methods focused on the flow of blood and clot formation and then we will talk about the methods to study the mechanics of such formed clots.

**System Level Methods**

System level methods are modeling of the blood flow and coagulation from a bigger scale, more of a mathematical relationship to reproduce the experimental results. These are generally modeled using CDR equations and the constants are estimated empirically from the experiments. The advantage of these types of models are with less complexity we can simulate the qualitative behavior of events related to the blood clotting. At the same time due to the assumptions made, the finer details of the processes are lost. A simple example of this type of method is the simulation of fibrin clotting time using simple ordinary differential equations[42] (ODEs). Blood coagulation factors are modeled as time dependent concentrations and chemical reactions are simulated using simple ODEs. Finally, clotting time of fibrin from fibrinogen in the presence of thrombin is simulated and matched with the experiments.

An elegant way to simulate enzyme reaction kinetics and associated products is Michaelis-Menten kinetics[43], which relates reaction rate to the substrate concentration. This method enables to define the enzymatic product formation using simple relationship like $E + S \xrightleftharpoons{k_f, k_r} ES \xrightarrow{k_{cat}} E + P$. Here, an enzyme ($E$) binding to a substrate ($S$) to form a complex ($ES$), which eventually releases a product ($P$). They reacts at a forward rate of $k_f$, reverse rate of $k_r$ and a catalytic rate of $k_{cat}$ and are collectively called as reaction rate constants or simply rate constants[44]. These equations can be written as a system of nonlinear ODEs[45] that define the rate of change of reactants. Using this technique, blood clot formation can be simulated[46], wherein the blood factors involved in the coagulation cascade is modeled as a system of Michaelis–Menten equations. Then based on the physiologically observed concentrations and rate constants, the kinetics of the reactions are simulated. One of the drawbacks of these methods is the dependency on empirical models and experimental values for rate constants, which makes the method susceptible of incorrect kinetics. In order to utilize the power of this method, one has to thoroughly validate the rate constants with experimental results.

Weisel [11] developed a kinetic model which can simulate the formation of the fibrinogen clot from individual monomers. They modeled the various stages of fibrin polymerization using simple rate equations. To account for the observed lag period seen in experiments, they have modified the model with additional fibrin geometric parameters. One drawback as they point out is the lack of accurate rate constants for many reactions considered, so they are chosen arbitrarily and later adjusted to accommodate the experimental results. However, a qualitative reasoning of the dependence of concentration and other factors on clotting rate is provided.

The system level reaction kinetics can be combined with spatial information to make it more realistic. In one such study, researchers combined the NS equations and CDR equations to study the blood clot formation and lysis[47]. The enzymatic reactions are described using 25 CDR equations and coupled with NS equations to account for the flow. The initiation and lysis of the clot formation is defined using threshold limits of fibrinogen concentrations. To initiate the CDR equations, platelet activation conditions are employed using an exponential function with a complex set of conditions.

**Continuum Methods**

While the system level methods can point out to the rate at which the clotting or lysis occurring in a particular hemostatic event, they fall short in understanding of clotting and lysis at physical length scales and geometries. This becomes important when studying the blood clots in complex geometries[48] and under various mechanical and environmental factors. Figure 3a represents a schematic of simulation of blood clot formation using continuum methods. In its simplest form, blood is modeled as an incompressible Newtonian fluid, flowing between two parallel flat plates. This will lead to a Poiseuille flow and the velocity across the flow cross section takes a parabolic form[29]. The flow profile will be disrupted when thrombus (semi solid formation) is formed near to the walls as shown in Fig. 3b. The complexity of the models depends on the level of details of the blood constituents considered. A basic continuum model is a combination of NS equations for flow dynamics and CDR equations for species spatiotemporal concentration evolution [49-51]. Most of the available continuum models assume blood vessels as rigid walls and blood constituents as massless particles. This enables to model the equations just by forward coupling NS equations' solution to CDR equations and neglecting the effect of particles on back on flow velocity.

Fogelson [52] modeled blood in a two-dimensional (2-D) domain as an incompressible fluid, and the flow is simulated using NS equation. The transport of platelets and adenosine diphosphate (ADP) in the blood stream is modeled using a particulate CDR equation. The solution of the NS equation for the Poiseuille flow is determined and then coupled with the CDR equations to estimate the impact of flow velocity on ADP and platelets. The mesh point velocities are interpolated to the particle positions using 2-D delta function. The activation of platelets, the release of ADP by them are simulated using threshold values of ADP in contact. The platelets interact with each other using a force function which can make it adhered to injured wall, sticky among themselves, or simply in an inactivated state. The study captured the aggregation of the platelets on an injured site.

These types of studies are further extended by Govindarajan [53] to accommodate the thrombus formation at injury site. The flow is simulated using NS equations, and spatiotemporal platelet concentrations using CDR equations. The platelets and other factors are modeled as continuum entities with varying concentrations, and the thrombus growth is simulated comparing the increased local concentration of fibrin. The effect of thus formed thrombus on blood flow is simulated by including an additional porous medium term to the NS equations, which increases the viscosity locally at regions of high fibrin concentrations and such methods are called two-way coupled modeling.

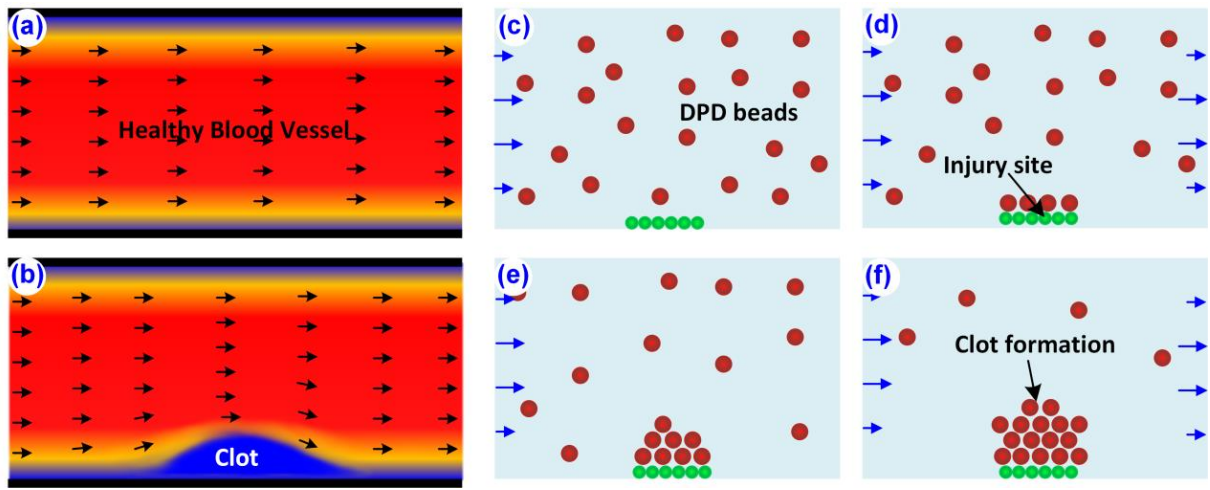

Figure 3. Schematic of continuum and DPD modeling of blood clot. Solution of NS equations for (a) a normal blood vessel and (b) blood vessel with thrombus formation. (c-f) Blood constituents, walls and plasma are modeled using DPD particles and used to simulate the rule based clot formation. Often DPD simulations are coupled with NS and CDR equations.

There are many studies aimed at applying the continuum level methods to specific cases of blood clots namely the aortic dissection [54-55], cerebral aneurysms [56-57] etc. The studies on these complex geometries can be performed by defining the NS equations for flow with appropriate boundary conditions like pulsating flow etc. and two-way coupling of the NS equations with CDR equations using momentum source term in NS equation. This will treat the growing thrombus as a region with high resistance to flow and hence simulate a solid obstruction (thrombus) to the flow.

So far we have seen the modeling of thrombus formation as an increase in concentration of platelets, fibrin or another blood constituent. This can be also modeled as a volume fraction of platelet deposition at the thrombus formation region [58]. Their study also accounted for a large number of blood constituents and their reactions through CDR equations. The activation and deposition of the platelets, thrombus propagation, erosion, inhibition, stabilization and thrombus-fluid interactions are captured in this study.

**Discrete Particles Methods**

Due to its versatile fluid simulation capabilities at both nano and micro scales, DPD method has caught attention of researchers to model blood flow[59-63]. The platelets are usually modeled as DPD particles or beads and fibrin as polymer chain beads connecting each other by harmonic springs. A typical particle based or DPD simulation setup for blood flow will look as shown in Fig. 3c-f.

When only platelets are considered for simulation, the DPD system constitutes only loosely distributed beads interacting with each other. Pressure or velocity boundary conditions are often applied to the system via external forces. These external forces can be solution of a simple Poiseuille flow or by coupling with NS equations. Filipovic[59] used CDR equations to estimate the local concentrations of blood constituents and NS equations to solve the velocity field. The walls are considered to be rigid and a velocity bounce back criteria was used to simulate the interaction between beads and wall. In this study NS equations are solved to get the flow field as an input to CDR and for DPD beads, a known solution of the Poiseuille flow (parabolic profile) velocity is given as the inlet and outlet boundary conditions.

One challenge with the DPD simulation is the modeling of dynamic bead adhesion to the wall (thrombus formation). Filipovic[59] used an additional force based model to capture this: $F_w^a = k_{bw}(1 - L_w/L_{max}^{wall})$. Here, $F_w^a$ is the attractive force between a bead and wall, $L_w$ is the distance of the bead from the wall and $L_{max}^{wall}$ is the size of the domain. The density distribution of blood factors obtained from the CDR solution is compared with the experimental results and the effective spring constant $k_{bw}$ is iteratively modified to match the results. This method is modified by Tosenberger[60], to include time dependent bead adhesion force. Their study has shown the formed thrombus ruptures at higher flow due to induced forces. Both of these studies didn't account for the blood constituents like fibrinogen and other important factors. This model is improved by adding CDR equations for fibrin concentration[61], thrombin concentrations [62], and basic coagulation pathway model using CDR equations [63].

**Multiscale Methods**

In multi-scale methods, two or more computational methods are coupled together at different length or time scales, which is different from hybrid methods which couple two or more methods at the same scales. A typical scenario of multi-scale modeling as shown in Fig. 4, is the modeling of blood factors' biochemical reactions using MD simulations, their spatiotemporal concentrations through CDR equations and blood flow using NS equations. This way, one can capture the macroscopic behavior of the clot formation without losing the nanoscale characteristics.

Xu et al.[64] developed a 2-D multi-scale model with fluid flow simulated using NS equations, spatiotemporal thrombin concentration using CDR equations, and a discrete CPM model for platelet activation, migration and binding behavior. The fibrin concentration in the system is modeled using a rate equation related to the concentration of the activated platelet. The simulation results are compared with the thrombus growth rates and sizes. This model is added with more blood cells and used Lagrangian coherent structure[65] (LCS) to capture more details of the blood clot formation[66]. The NS equations are coupled with the CPM model to capture the effect of cell adhesion on blood flow and vice versa. However, smaller particles (cells) are not coupled with NS to reduce the computational expense. The model is continually improved by adding CDR equations to accommodate the surface mediated clot formation based on the coagulation cascade[67]. As a variation, fibrin polymerization is modeled by introducing additional differential equations, fibrin element sub model[68], which also extends a 3-D version of LBM coupled with sub cellular element model (SCEM) to simulate blood flow.

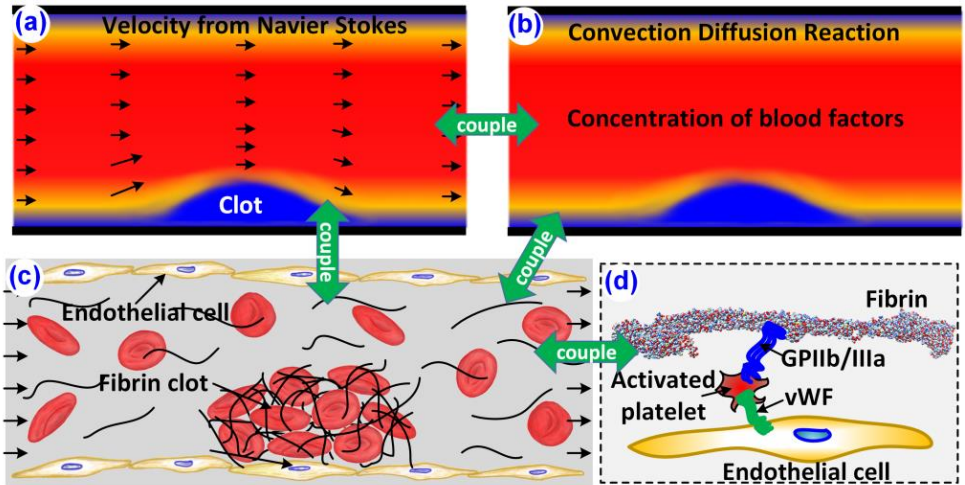

Figure 4. Schematic of multiscale blood clot modeling approach. (a) In continuum length and time scale NS equations are solved to get velocity field. (b) CDR equations give the solutions as the concentration of various blood factors and their reactions. (c) Clot formation using mesoscopic methods like DPD, CGMD or CPM. (d) MD simulations or Ab-Initio methods to capture atomic level biomolecular interactions.

Instead of using CPM and CDR equations, Moreno[69] formulated the multi-scale blood flow using a coupled NS equation, variational multi-scale method and smooth dissipative particle dynamics[70] (SDPD). However, this method needs a lot of improvement, like including blood factors and other criteria. A brief review about the particle based modeling methods for blood flow, application of Stokesian dynamics with a discrete element method, Particle semi-implicit method and DPD is discussed by Yamaguchi[10].

**Modeling of Clot Mechanics**

In the previous sections we have discussed the methods and models used to simulate the flowing blood and formation of thrombus and here we will focus on the models which is used to study the mechanical properties of fibrin or blood clot. The study of structure and mechanics behind the thrombus is of great importance since it can predict the physiological or pathological conditions at which it can break, leading to VTE or other states. A brief review of interfacial mechanics of biomaterials is available in literature[71]. Many experimental investigations and empirical models are aimed to predict the compression-decompression behavior of the blood clots[72-73]. There exist modular models to simulate the stress strain behavior of fibrin fibers[6], which can be scaled up and combined with other models to predict the bulk behavior of fibrin clots.

# Discussion and Future Trends

The trend of the computational modeling over the years show that the simpler models like single component continuum models are becoming redundant and increasingly replaced with discrete particle methods like DPD and with the emerging technique called multiscale modeling. The continuum level methods are still suitable for predicting the top level behavior based on empirical data, limiting them to use NS equations coupled with CDR equations and investigation of the blood flow dynamics. Two-way coupling of NS equations with CDR or other methods is superior to one-way coupling in many aspects like accurate flow distribution of blood, thrombus deformations due to high shear flow etc.

DPD is a powerful method for simulating fluids, which nowadays find place in many biological simulations. The base requirements for using DPD in blood flow simulations are put forward by many[10, 32, 59, 61, 63] and is continually improving. One of the advantage of DPD is that we can model many of the complex blood components like WBCs, RBCs, platelets and plasma as simple beads. These properties make DPD as a better choice compared to boundary element methods, LBM, discrete element method etc. One of the biggest challenge that the field still experiences is the modeling of blood as a non-Newtonian fluid, where DPD shows some promise[74-76]. However, one has to be careful in introducing additional potentials in DPD to simulate polymer chains, solid walls, dynamic bond formation and breaking etc., since DPD was originally developed for fluids with only three main soft interactive forces. Any modification on them will alter the dynamics and the parameters has to be validated with experimental values.

Multiscale blood clot models as shown in Fig. 4, can simulate the blood clotting dynamics, biochemical reactions, factor concentrations etc. effectively. The model will have cascading submodels which can serve as independent models to predict a few specific characteristics like thrombin cleaving fibrinogen, attachment of fibrin to vWF etc. to submodels which can predict thrombus solid liquid interactions etc. A good multiscale model will have the characteristics of simulating the bulk behavior of the blood like diffusivity, viscosity etc., mechanical properties of the thrombus under shear flow, and dynamic properties to simulate lysis and thrombus rupture based on the molecular level mechanics. With the development of more accurate models for subcomponents of blood constituents, it will be promising to develop and simulate realistic blood clot formation and lysis. Erythrocyte multiscale models[77-78] can predict the behavior of RBCs in the plasma flow, fibrin submodels[79-81] can simulate atomic level[82] to mesoscale properties. The mathematical modeling of the overall coagulation process is advancing with capturing more and more factors and conditions[83-87]. These advancements in different scales will enable us to develop new accurate multiscale models to predict both blood clotting process and its mechanics. In table 1, some of the strength and weakness of different scales of modeling is provided.

**Table 1**. Strength and Weakness of Various Methods

| Method | Advantage | Weakness |
|---|---|---|
| System level | • Good agreement with experiments | • Missing spatiotemporal information |
| Continuum | • Easy to setup<br>• Less computational power | • Molecular and microscopic details are missing<br>• Need CDR equations to account for blood factors |
| DPD | • Can work with complex geometries | • Computationally expensive for big systems |
| Multiscale | • Detailed information of system included<br>• Fine-tuned information at any scale can be included to improve the modeling<br>• Reliable simulation results of system under consideration | • Complex to model<br>• Needs careful designing of algorithms to solve at different scales<br>• Computationally expensive |

## Conclusion

In this paper we have reviewed the recent developments in the area of computational modeling of blood clot formation using various methods at different length and time scales. From continuum level methods like NS equation coupled with CDR equations and mesoscopic methods like CPM, DPD and SPH are reviewed. A promising newly developed class of multiscale methods which utilizes the power of methods at different length and time scales is discussed in detail. The trend and the future direction leads to more multiscale methods to be evolved. We have also critically analyzed some of the strength and weakness of the methods and suggested the better ones.

## Acknowledgements

Research reported in this publication was supported by the National Heart, Lung, and Blood Institute of the National Institutes of Health under Award Number 1K01HL115486. The content is solely the responsibility of the authors and does not necessarily represent the official views of the National Institutes of Health.

## Nomenclature

| | |
|---|---|
| I | Fibrinogen |
| II | Prothrombin |
| III | Tissue Thromboplastin |
| IV | Calcium |
| V | Proaccelerin |
| VII | Proconvertin |
| VIII | Antihemophilic factor A |
| IX | Antihemophilic factor B |
| X | Thrombokinase |
| XI | Plasma thromboplastin antecedent |

| XII | Hageman factor |
| XIII | Fibrin-stabilizing factor |
| MD | Molecular Dynamics |